\documentstyle[11pt,amssymb,epsf]{article}

\def\ben{\begin{equation}}
\def\een{\end{equation}}

\let\a=\alpha    
    
  \let\n=\nu

\let\C=\Chi

\def\nn{\nonumber} \def\bd{\begin{document}} \def\ed{\end{document}}
\def\ds{\documentstyle} \let\fr=\frac \let\bl=\bigl \let\br=\bigr
\let\Br=\Bigr \let\Bl=\Bigl
\let\bm=\bibitem
\let\na=\nabla
\let\pa=\partial \let\ov=\overline
\newcommand{\be}{\begin{equation}}
\newcommand{\ee}{\end{equation}}
\def\ba{\begin{array}}
\def\ea{\end{array}}
\def\ft#1#2{{\textstyle{{\scriptstyle #1}\over {\scriptstyle #2}}}}
\def\fft#1#2{{#1 \over #2}}
\def\del{\partial}
\def\vp{\varphi}
\def\sst#1{{\scriptscriptstyle #1}}
\def\oneone{\rlap 1\mkern4mu{\rm l}}
\def\td{\tilde}
\def\wtd{\widetilde}
\def\ie{\rm i.e.\ }
\def\dalemb#1#2{{\vbox{\hrule height .#2pt
        \hbox{\vrule width.#2pt height#1pt \kern#1pt
                \vrule width.#2pt}
        \hrule height.#2pt}}}
\def\square{\mathord{\dalemb{6.8}{7}\hbox{\hskip1pt}}}
\newcommand{\ho}[1]{$\, ^{#1}$}
\newcommand{\hoch}[1]{$\, ^{#1}$}
\newcommand{\bea}{\begin{eqnarray}}
\newcommand{\eea}{\end{eqnarray}}
\newcommand{\ra}{\rightarrow}
\newcommand{\lra}{\longrightarrow}
\newcommand{\Lra}{\Leftrightarrow}
\newcommand{\ap}{\alpha^\prime}
\newcommand{\bp}{\tilde \beta^\prime}
\newcommand{\tr}{{\rm tr} }
\newcommand{\Tr}{{\rm Tr} }
\def\0{{\sst{(0)}}}
\def\1{{\sst{(1)}}}
\def\2{{\sst{(2)}}}
\def\3{{\sst{(3)}}}
\def\4{{\sst{(4)}}}
\def\5{{\sst{(5)}}}
\def\6{{\sst{(6)}}}
\def\7{{\sst{(7)}}}
\def\8{{\sst{(8)}}}
\def\n{{\sst{(n)}}}
\def\cA{{{\cal A}}}
\def\cF{{{\cal F}}}
\def\tV{\widetilde V}
\def\tW{\widetilde W}
\def\tH{\widetilde H}
\def\tE{\widetilde E}
\def\tF{\widetilde F}
\def\tA{\widetilde A}
\def\im{{{\rm i}}}
\def\tY{{{\wtd Y}}}
\def\ep{{\epsilon}}
\def\vep{{\varepsilon}}
\def\R{\rlap{\rm I}\mkern3mu{\rm R}}
\def\bD{{{\bar D}}}

\def\R{\rlap{\rm I}\mkern3mu{\rm R}}
\def\bD{{{\bar D}}}
\def\R{{{\Bbb R}}}
\def\C{{{\Bbb C}}}
\def\H{{{\Bbb H}}}
\def\CP{{{\Bbb C}{\Bbb P}}}
\def\RP{{{\Bbb R}{\Bbb P}}}
\def\Z{{{\Bbb Z}}}
\def\bA{{{\Bbb A}}}
\def\bB{{{\Bbb B}}}
\def\bC{{{\Bbb C}}}
\def\bD{{{\Bbb D}}}
\def\bZ{{{\Bbb Z}}}
\def\bK{{{\Bbb K}}}
\def\Re{{{\frak{Re}}}}
\def\cosec{{\,\hbox{cosec}\,}}

\newcommand{\tamphys}{\it Center for Theoretical Physics,
Texas A\&M University, College Station, TX 77843, USA}
\newcommand{\umich}{\it Michigan Center for Theoretical Physics,
University of Michigan\\ Ann Arbor, MI 48109, USA}
\newcommand{\upenn}{\it Department of Physics and Astronomy,
University of Pennsylvania\\ Philadelphia,  PA 19104, USA}
\newcommand{\SISSA}{\it  SISSA-ISAS and INFN, Sezione di Trieste\\
Via Beirut 2-4, I-34013, Trieste, Italy}

\newcommand{\ihp}{\it Institut Henri Poincar\'e\\
  11 rue Pierre et Marie Curie, F 75231 Paris Cedex 05}

\newcommand{\damtp}{\it DAMTP, Centre for Mathematical Sciences,
 Cambridge University\\ Wilberforce Road, Cambridge CB3 OWA, UK}
\newcommand{\itp}{\it Institute for Theoretical Physics, University of
California\\ Santa Barbara, CA 93106, USA}

\newcommand{\auth}{M. Cveti\v{c}\hoch{\dagger}, G.W. Gibbons\hoch{\sharp},
H. L\"u\hoch{\star} and C.N. Pope\hoch{\ddagger}}

\begin{document}
\begin{flushright}
\hfill{DAMTP-2001-110}\ \ \ {CTP TAMU-35/01}\ \ \ {UPR-974-T}\ \ \
{MCTP-01-63}\\
{December 2001}\\
{hep-th/0112138}
\end{flushright}


\begin{center}
{ \large {\bf A $G_2$ Unification of the Deformed and Resolved Conifolds}}

\vspace{5pt}
\auth

\vspace{3pt}
{\hoch{\dagger}\upenn}

\vspace{3pt}


\vspace{3pt}
{\hoch{\sharp}\damtp}

\vspace{3pt}
{\hoch{\star}\umich}

\vspace{3pt}
{\hoch{\ddagger}\tamphys}

\vspace{3pt}

\underline{ABSTRACT}
\end{center}

    We find general first-order equations for $G_2$ metrics of
cohomogeneity one with $S^3\times S^3$ principal orbits.  These reduce
in two special cases to previously-known systems of first-order
equations that describe regular asymptotically locally conical (ALC)
metrics $\bB_7$ and $\bD_7$, which have weak-coupling limits that are
$S^1$ times the deformed conifold and the resolved conifold
respectively.  Our more general first-order equations provide a
supersymmetric unification of the two Calabi-Yau manifolds, since the
metrics $\bB_7$ and $\bD_7$ arise as solutions of the {\it same}
system of first-order equations, with different values of certain
integration constants.  Additionally, we find a new class of ALC $G_2$
solutions to these first-order equations, which we denote by
$\wtd\bC_7$, whose topology is an $\R^2$ bundle over $T^{1,1}$.  There
are two non-trivial parameters characterising the homogeneous
squashing of the $T^{1,1}$ bolt.  Like the previous examples of the
$\bB_7$ and $\bD_7$ ALC metrics, here too there is a $U(1)$ isometry
for which the circle has everywhere finite and non-zero length.  The
weak-coupling limit of the $\wtd\bC_7$ metrics gives $S^1$ times a
family of Calabi-Yau metrics on a complex line bundle over $S^2\times
S^2$, with an adjustable parameter characterising the relative sizes
of the two $S^2$ factors.

{\vfill\leftline{}\vfill
\vskip 5pt
\footnoterule
{\footnotesize \hoch{\dagger} Research supported in part by DOE grant
DE-FG02-95ER40893 and NATO grant 976951. \vskip -12pt} \vskip 14pt
{\footnotesize \hoch{\star} Research supported in full by DOE grant
DE-FG02-95ER40899. \vskip -12pt} \vskip 14pt
{\footnotesize  \hoch{\ddagger} Research supported in part by DOE
grant DE-FG03-95ER40917.\vskip  -12pt}}

\pagebreak
\setcounter{page}{1}

\vfill\eject

\section{Introduction}

   Seven-dimensional manifolds of $G_2$ holonomy are of considerable
interest in string theory and M-theory, since they are closely related
\cite{atmava,achvaf,atiwit} to the six-dimensional Calabi-Yau
manifolds that have formed the basis of most attempts to extract
phenomenologically realistic four-dimensional physics from string
theory.  Although the principal phenomenological focus would be on
compact internal manifolds, implying a discrete spectrum of massive
and massless four-dimensional fields, it is often useful to study
non-compact manifolds, since these can form the ``building blocks''
that describe the regions of localised curvature near an orbifold
limit of a smooth compact space.
 
   If a non-compact $G_2$ metric has an asymptotically locally conical
(ALC) geometry, meaning that at large distance it approaches a twisted
product of a circle and a six-dimensional asymptotically conical (AC)
metric, then this circle may be used in a Kaluza-Klein reduction that
allows a reinterpretation of an M-theory solution as a type IIA string
solution.  If there is a non-trivial parameter in the $G_2$ metric
that allows one to adjust the asymptotic radius of the circle, while
holding the scale-size of the interior ``core'' fixed, then one can
take a limit of vanishingly-small radius for the circle that therefore
corresponds to weak coupling in the type IIA string picture.  In this
weak-coupling limit, known mathematically as the ``Gromov-Hausdorff
limit,'' the remaining six-dimensional metric becomes exactly a
Ricci-flat K\"ahler metric; \ie a Calabi-Yau metric, with $SU(3)$
holonomy.

    The non-compact Calabi-Yau metrics that have received the most
attention are those associated with the singular conifold, and its
smoothed-out versions.  The conifold itself is the Ricci-flat metric
on the cone over the homogeneous Einstein metric on $T^{1,1}
=(S^3\times S^3)/S^1$, which becomes singular at the origin where the
radius of the $T^{1,1}$ collapses to zero.  It can be smoothed out in
two different ways, with the apex of the cone either becoming a smooth
2-sphere or a smooth 3-sphere \cite{candel}.  The former case is
called the resolved conifold and the latter is called the deformed
conifold.  

    An ansatz for a class of cohomogeneity one $G_2$ metrics with
$S^3\times S^3$ principal orbits was introduced \cite{bragogugu}.  The
ansatz has four metric functions, and a system of four first-order
equations was derived in \cite{bragogugu}, by requiring that the
metrics have $G_2$ holonomy.  These equations admit solutions
describing ALC metrics with a degenerate $S^3$ orbit (an $S^3$ bolt)
at short distance.  A specific solution was obtained in
\cite{bragogugu}, and it was argued that this was but one member of a
one-parameter family of ALC metrics.  In \cite{cglp2}, the general
regular short-distance Taylor expansions for solutions with an $S^3$
bolt were obtained, which indeed contained one non-trivial adjustable
parameter.  By numerical integration of the first-order equations
governing the $G_2$ holonomy, it was shown that within a certain range
for this non-trivial parameter, the metrics are ALC and regular also
at large distance.  The adjustable parameter in this family of
solutions, which were denoted by $\bB_7$ in \cite{cglp2},
characterises the asymptotic radius of the circle factor, while the
radius of the $S^3$ bolt is held fixed.  At the upper end of the
parameter range the radius becomes infinite, and the $G_2$ metric
becomes the original AC metric found in \cite{brysal,gibpagpop}.  At
the lower end of the parameter range, the asymptotic radius of the
circle goes to zero, and the metric approaches $S^1$ times the
Ricci-flat K\"ahler metric on the deformed conifold.  An important
feature of the one-parameter family of $\bB_7$ metrics is that the
$S^3$ bolt is always ``round'' (with its Einstein metric), regardless
of the value of the parameter.

   A more general class of metric ansatz, again with cohomogeneity one
and $S^3\times S^3$ principal orbits, was introduced in \cite{cglp2}.
There are nine functions in the metric ansatz, and the isometry group
is $SU(2)\times SU(2)$ acting by left translations on the orbits.  In
order to be able to make a Kaluza-Klein relation to Calabi-Yau
metrics, a specialisation of this to a six-function ansatz was made
recently in \cite{cglpcon}.  The restriction from nine to six
functions, the metric ansatz appearing in (\ref{d7ans0}) below, is
chosen so as to give an additional right-acting diagonal $U(1)$
isometry.  This $U(1)$ isometry plays an important role in the
subsequent Kaluza-Klein reduction.  By then writing down a natural
choice of associative 3-form, and imposing closure and co-closure (the
necessary and sufficient conditions for $G_2$ holonomy), a new system
of first-order equations for $G_2$ holonomy was obtained in
\cite{cglpcon}.\footnote{An apparently equivalent system, but with a
second-order equation, has since appeared in \cite{brand}.}  This
system comprised four independent first-order equations, together with
two algebraic expressions for the remaining two metric functions.
(The previous system of first-order equations found in
\cite{bragogugu} can also be embedded within the more general class
considered in \cite{cglpcon}, as an {\it inequivalent} set of four
independent first-order equations with two algebraic expressions for
the remaining two metric functions.)  The new system of four
first-order equations obtained in \cite{cglpcon} was shown to admit
short-distance Taylor expansions describing regular metrics with an
$S^3$ bolt.  This time, however, there is a non-trivial adjustable
parameter that characterises the degree of homogeneous ``squashing''
of the $S^3$ bolt.  It was then shown in \cite{cglpcon}, by using
numerical integration methods, that for a suitable range of the
parameter one gets ALC metrics, denoted by $\bD_7$, that are regular
also at large distance.  The upper end of the parameter range, as the
radius of the stabilising circle goes to infinity, corresponds again
to the case of the AC metric found in \cite{brysal,gibpagpop}.  At the
lower end of the parameter range, \ie in the Gromov-Hausdorff limit,
the $S^3$ bolt becomes infinitely squashed to $S^2$, and one obtains
$S^1$ times the resolved conifold \cite{cglpcon}.

   With these two examples, the $\bB_7$ and the $\bD_7$ metrics, one
has the ability to obtain both the deformed conifold and the resolved
conifold as weak-coupling limits of families of seven-dimensional $G_2$
metrics.  In fact since both $\bB_7$ and $\bD_7$ have a coincident 
AC limit at the upper ends of their parameter ranges, one could say
that deformed and the resolved conifold solutions in weakly-coupled
string theory are related via a strong-coupling regime and the
eleven-dimensional M-theory \cite{cglpcon}.

   The two systems of four first-order equations that give rise to the
$\bB_7$ and $\bD_7$ families of $G_2$ metrics both, of course, imply
that the conditions of Ricci-flatness are satisfied.  Thus the $\bB_7$
and $\bD_7$ metrics come from a common six-function metric ansatz and
they satisfy the {\it same} system of Ricci-flat equations, with
certain specific (consistent) truncations implied by the pair of
algebraic constraints that were imposed in the two cases.  Thus the
statement in \cite{cglpcon} about the strong-coupling relation between
the deformed and the resolved conifolds was at the level of solutions
of the M-theory field equations.  It could not be argued in
\cite{cglpcon} that there was a connecting path within the stronger
criterion of $G_2$ holonomy, since the two systems of first-order
equations for the two cases were ostensibly disconnected.

   In order to be able to link the deformed and the resolved conifolds
via a $G_2$ (and hence supersymmetric) path, we need to find a larger
system of first-order equations, within the framework of the
six-function metric ansatz, which not only ensures $G_2$ holonomy but
also encompasses both of the previous four-equation first-order
systems.  The search for such an enlarged $G_2$ system forms the
subject of this paper.  We shall show that the most general system of
equations determining $G_2$ holonomy, starting from our six-function
metric ansatz, is a system comprising five independent 
first-order equations, together with one algebraic expression for
the sixth metric function.  Both of the previous systems of 
four first-order equations are then contained as
(consistently truncated) special cases of our more general
equations. Thus, in particular, we now have a supersymmetric path of
$G_2$ metrics relating the deformed and the resolved conifolds.

   An additional outcome from our new $G_2$ equations is that we can
also fit some further six-dimensional Calabi-Yau metrics into the
picture.  Long ago, a general construction of Ricci-flat K\"ahler
metrics on line bundles over Einstein-K\"ahler bases spaces was found
\cite{berber,pagpop}.  A particular example is the six-dimensional
case of the line bundle over $S^2\times S^2$.  This is a metric of
cohomogeneity one, with principal orbits that are $T^{1,1}/Z_2$.  In
fact this particular metric was shown in \cite{slumpy} to arise as the
Gromov-Hausdorff limit of another class of regular ALC $G_2$ metrics
that were obtained there.  They are solutions of the four-function
system introduced in \cite{bragogugu}.  In these metrics, denoted by
$\bC_7$, there is a bolt that is a squashed $T^{1,1}$, with a
non-trivial parameter characterising the length of the $U(1)$ fibres
relative to the Einstein $S^2\times S^2$ base.  At the upper limit of
the parameter range the metrics are AC, and asymptotically approach
the AC $G_2$ metric of \cite{brysal,gibpagpop}.  At the lower end of
the parameter range the length of the $U(1)$ fibres goes to zero, and
one obtains $S^1$ times the metric \cite{berber,pagpop} on the complex
line bundle over the Einstein $S^2\times S^2$.

  More general Ricci-flat K\"ahler metrics on the complex line bundle
over $S^2\times S^2$ can also be obtained, in which the sizes of the
two $S^2$ factors can be adjusted arbitrarily \cite{pando2}.  Until
now, it has not been possible to obtain these as a weak-coupling limit
of any known $G_2$ metrics.  It turns out that the more general system
of $G_2$ metrics with five first-order equations that we obtain in
this paper admit further regular ALC solutions, which we shall denote
by $\wtd\bC_7$, whose Gromov-Hausdorff limits are precisely the more
general Ricci-flat K\"ahler metrics of \cite{pando2}.  As in the
$\bC_7$ and $\bD_7$ metrics that we studied previously, again here we
find that the length of the circle associated with the $U(1)$ 
isometry in the $\wtd\bC_7$ metrics is everywhere finite and
non-zero.  This allows a non-singular Kaluza-Klein reduction with a 
nowhere singular dilaton in the type IIA picture.

\section{The general $G_2$ metrics with $S^3\times S^3$ 
   principal orbits}

\subsection{The metric ansatz, and first-order equations}\label{metfosec}

   Our starting point is an ansatz for seven-dimensional metrics of
cohomogeneity one, and $S^3\times S^3$ principal orbits, 
that was used in \cite{cglpcon}:
\be
ds_7^2 = dt^2 + a^2\, [(\Sigma_1+ g\, \sigma_1)^2 +
(\Sigma_2+ g\,\sigma_2)^2] + b^2\, (\sigma_1^2 + \sigma_2^2) +
c^2 (\Sigma_3 +g_3\, \sigma_3)^2 + f^2 \sigma_3^2\,,\label{d7ans0}
\ee
In fact, it turns out that the subsequent equations for $G_2$ 
holonomy are greatly simplified by working with a different set of
variables $(\td c,\td f\, \td g_3)$, in terms of which (\ref{d7ans0}) 
becomes
\be
ds_7^2 = dt^2 + a^2\, ((\Sigma_1+ g\, \sigma_1)^2 +
(\Sigma_2+ g\,\sigma_2)^2) + b^2\, (\sigma_1^2 + \sigma_2^2) +
\td c^2 (\Sigma_3 -\sigma_3)^2 + \td f^2 \,(\Sigma_3+ 
   \td g_3\, \sigma_3)^2
\,,\label{d7ans}
\ee
where $a$, $b$, $\td c$, $\td f$, $g$ and $\td g_3$ are functions only of the
radial variable $t$, and $\sigma_i$ and $\Sigma_i$ are left-invariant
1-forms of $SU(2)\times SU(2)$.  They can be expressed in terms of
Euler angles as 
\bea
&&\sigma_1+\im\, \sigma_2= e^{-\im\, \psi}\, (d\theta + \im\,
\sin\theta\, d\phi)\,,\qquad \sigma_3=d\psi + \cos\theta\,
d\phi\,,\nn\\
&&\Sigma_1+\im\, \Sigma_2= e^{-\im\, \wtd\psi}\, (d\td\theta + \im\,
\sin\td\theta\, d\td\phi)\,,\qquad \Sigma_3=d\wtd\psi + \cos\td\theta\,
d\td\phi\,.
\eea
The metric is a specialisation of a nine-function ansatz introduced in
\cite{cglp2}, in which all three directions $\sigma_i$ and $\Sigma_i$
were given distinct metric functions.  The nine-function ansatz has the
left-acting $SU(2)\times SU(2)$ group as isometries.  As discussed in
\cite{cglpcon}, by setting the $i=1$ and $i=2$ directions equal as in
(\ref{d7ans}), we gain an additional diagonal $U(1)$ of right-acting
isometries, generated by the Killing vector
\be
K= \fft{\del}{\del\psi} + \fft{\del}{\del\wtd\psi}\,.\label{kvec}
\ee

   In \cite{cglpcon}, we obtained first-order equations for $G_2$
holonomy, by writing down a specific ansatz for an associative 3-form.
Here, we shall follow a slightly different strategy, and instead
demand that there exist a covariantly-constant (\ie parallel) spinor
$\eta$.  Having a covariantly-constant spinor is equivalent to having
a closed and co-closed associative 3-form; they both imply $G_2$
holonomy.  However, our result that we shall get from using the
requirement of a covariantly-constant spinor is more general than the
result we obtained in \cite{cglpcon}, since the associative 3-form in
\cite{cglpcon} was not the most general that is allowed by the
isometries of the metric, whilst we make no analogous restrictive
assumption for the covariantly-constant spinor.  In consequence we
shall obtain the most general possible set of first-order equations
giving $G_2$ holonomy for the metric ansatz (\ref{d7ans}).

   After imposing the requirement on (\ref{d7ans}) that
there exist a covariant-constant spinor $\eta$,  $D\,
\eta\equiv d\, \eta + \ft14\omega_{ab}\, \Gamma^{ab}\,\eta=0$, we
obtain an algebraic expression for $\td g_3$,
\be
\td g_3 = \fft{ a^3\, \td c\, g + (b\, \td f- 
   a\, \td c\, g)\, W^2}{a^2\, b\, \td f}\,,\label{constr}
\ee
together with first-order equations for the five remaining metric 
functions,
\bea
a' &=& \fft{a^4\, g^2 - \td c^2\, W^2}{2 a\, b\, \td c\, W}\,,\nn\\
b' &=& \fft{a^5\, g^4 - 
[a\, \td c^2 + 2b\, \td c\, \td f\, g + a\, 
       (a^2-\td c^2)\, g^2]\, W^2}{2 a\, b^2\, \td c\, W}\,,\nn\\
{\td c}' &=& \fft{ a^2\, \td c^2 + 
   (\td c^2-2 a^2)\, W^2}{2 a^2\, b\, W}\,,
\label{5fo}\\
{\td f}' &=& \fft{a^2\, \td c\, g\, (a\, \td f\, g- 2b\, \td c)
- g\, [a\, \td c\, \td f\, g 
   -2b\, (\td c^2+\td f^2)]\, W^2}{2 a\, b^3\, W}\,,\nn\\
g' &=& \fft{a^3\, b\, g + \td c\, \td f\, W^2}{a^3\, \td c\, W}\,,\nn
\eea
where
\be
W\equiv \sqrt{b^2+a^2\, g^2}\,,\label{wdef}
\ee
and a prime denotes a derivative with respect to $t$.
One can straightforwardly verify, using the results for the Ricci
tensor obtained in \cite{cglp2}, that the metric (\ref{d7ans}) is
Ricci-flat if (\ref{constr}) and (\ref{5fo}) are satisfied.  

\subsection{Parallel spinor and calibrating 3-form}

  It is quite easy to give an explicit expression for the
covariantly-constant spinor that we found in the analysis in section
\ref{metfosec}.  We take the vielbein for the metric (\ref{d7ans}) to
be 
\bea
&&e^0=dt\,,\quad  e^1=a\, (\Sigma_1 +g\, \sigma_1)\,,\quad
e^2=a\, (\Sigma_2 +g\, \sigma_2)\,, \quad
e^3=\td c\, (\Sigma_3 -\sigma_3)\,,\nn\\
&&e^4=b\, \sigma_1\,,\quad
e^5=b\, \sigma_2\,,\quad  e^6=\td f\,(\Sigma_3 +\td g_3\, \sigma_3)\,.
\label{viel1}
\eea
Using the natural choice of spinor frame, we then find that the
covariantly-constant spinor $\eta$ is given by
\be
\eta=W^{-\ft12}\Big(\sqrt{b + \im\, a\, g}\, \epsilon_1 +
\sqrt{b-\im\, a\, g}\, \epsilon_2\Big)\,,
\ee
where $\epsilon_1$ and $\epsilon_2$ are spinors with constant components,
defined uniquely up to overall scale by 
\be
\Gamma_{14}\, \ep_1=\Gamma_{25}\, \ep_1 = \Gamma_{36}\, 
\ep_1 =\im\, \ep_1\,, \qquad 
\ep_2=\Gamma_{123}\, \ep_1 \,.\label{epcon} 
\ee
If we normalise $\ep_1$, and hence $\ep_2$, to unit length,
$\ep_1^\dagger\, \ep_1= \ep_2^\dagger\, \ep_2=1$, then we shall also
have $\eta^\dagger\, \eta=1$.

   From $\eta$, we can construct the closed and co-closed associative 
3-form $\Phi$, whose tangent-space components are given by $\Phi_{abc} 
=\im\, \eta^\dagger\, \Gamma_{abc}\, \eta$.  This turns out to be
\bea
\Phi &=& e^0\wedge (e^1\wedge e^4 + e^2\wedge e^5 + e^3\wedge e^6)
+(e^1\wedge e^2-e^4\wedge e^5)\wedge (\a\, e^3 + \beta\, e^6)\nn\\
&&+(e^1\wedge e^5-e^2\wedge e^4)\wedge (\beta\, e^3 - \a\, e^6)
\,,\label{3form}
\eea
where
\be
\a\equiv \fft{a\, g}{\sqrt{b^2 + a^2\, g^2}}\,,\qquad 
\beta\equiv \fft{b}{\sqrt{b^2+a^2\, g^2}}
\,.
\ee
Note that $\a^2+\beta^2=1$, and so the explicitly-appearing radial 
dependence in (\ref{3form}) is an $SO(2)$ rotation in the $(e^3,e^6)$ plane.
One can easily verify that $\Phi$ given in (\ref{3form}) is indeed 
a closed and co-closed associative 3-form.

\subsection{Two constants of the motion}\label{simpsec}

    Note that two combinations of the five first-order equations can
be integrated.  Specifically, if we define
\be
p \equiv -(a\, \td c\, g + b\, \td f\, \td g_3)\, W\,,\qquad
q \equiv \fft{a^2\, (a\, \td c\, g + b\, \td f)}{W} 
\,,\label{integrals}
\ee
where $\td g_3$ is given by
(\ref{constr}) and $W$ is given by (\ref{wdef}), then we shall have 
$dp/dt=0$ and $dq/dt=0$.  This can be seen by substituting the
first-order equations into these expressions, but a simpler way to see
the result is by noting that $p$ and $q$ are nothing but the coefficients
of the volume forms of the two 3-spheres 
in the expression (\ref{3form}) for the associative 3-form $\Phi$;
\be 
\Phi= p\, \sigma_1\wedge \sigma_2\wedge \sigma_3 + q\,
\Sigma_1\wedge\Sigma_2\wedge\Sigma_3 + \cdots\,.\label{pqapp}  
\ee
It is now evident, as observed in \cite{brand}, that the closure
condition $d\Phi=0$ implies that $p$ and $q$ must be constants.  If it
happens that $p=-q$, then the metric has a $Z_2$ symmetry under which
the two $S^3$ factors in the $S^3\times S^3$ principal orbits are
interchanged \cite{brand}.  When $p\ne -q$, the metric is not $Z_2$
symmetric.

    The two expressions (\ref{integrals}) could be used in order to
reduce the system of five first-order equations to a system with three
first-order equations and the two constants $p$ and $q$.  (A
formulation apparently equivalent to this, starting from a
$G_2$-invariant ansatz for $\Phi$ involving $p$ and $q$ as given
constants, and reducing to a second-order differential equation, was
obtained in \cite{brand}.  We understand that $G_2$ metrics with
constants equivalent to $p$ and $q$ have also been considered by
S. Gukov, K. Saraikin and A. Volovich \cite{gukov}.)

  It is very important, however, that in our complete system of five
first-order equations the quantities $p$ and $q$ arise as constants of
integration, rather than being fixed, given constants in an ansatz
with fewer independent metric functions.  In particular, it means that
solutions with different values of $p$ and $q$ can all be viewed as
solutions of the {\it same} system of five first-order equations that
imply $G_2$ holonomy.

\subsection{An alternative form for the metric}

    For some purposes, it is useful to make a further reorganisation 
of the metric on the orbit space, replacing (\ref{d7ans}) by the more
symmetrical ansatz 
\bea
ds_7^2 &=& dt^2 + \td a^2\, [(\Sigma_1 + \td g\, \sigma_1)^2 +
                          (\Sigma_2 + \td g\, \sigma_2)^2]
+ \td b^2\,  [(\Sigma_1 - \td g\, \sigma_1)^2 +
                          (\Sigma_2 - \td g\, \sigma_2)^2]\nn\\
&&+ \td c^2\, (\Sigma_3-\sigma_3)^2 + \td f^2\, (\Sigma_3 + \td g_3\,
\sigma_3)^2\,.\label{d7ans2}
\eea
After doing this, we find that the algebraic relation (\ref{constr}) 
becomes
\be
\td g_3 = \td g^2 - \fft{\td c\, (\td a^2-\td b^2)
                 (1-\td g^2)}{2\td a\, \td b\, \td f}\,,
\label{constr2}
\ee
while the first-order equations (\ref{5fo}) for the five remaining metric 
functions become
\bea
{\td a}' &=& \fft{\td c^2\, (\td a^2 -\td b^2) + 
[4\td a^2\, (\td a^2-\td b^2)- \td c^2\, (5 \td a^2-\td b^2) - 4\td
a\, \td b\, \td c\, \td f]\, \td g^2}{16\td a^2\, \td b\, \td c\, \td
g^2}\,,\nn\\
{\td b}' &=& -\, \fft{\td c^2\, (\td a^2-\td b^2) + [4 \td b^2\, 
(\td a^2 -\td b^2) +\td c^2\, (5\td b^2 - \td a^2) - 4\td a\, \td b\, 
\td c\, \td f]\,\td g^2}{16 \td a\, \td b^2\, \td c\, \td g^2}\,,\nn\\
{\td c}' &=& \fft{\td c^2 + (\td c^2 -2\td a^2 -2\td b^2)\, \td
g^2}{4\td a\, \td b\, \td g^2}\,,\label{5fo2}\\
{\td f}' &=& -\, \fft{(\td a^2-\td b^2)\, [ 4 \td a\, \td b\, \td f^2\,
\td g^2 - \td c\, (4\td a\, \td b\, \td c + \td a^2\, \td f - \td
b^2\, \td f)\, (1-\td g^2)]}{16 \td a^3\, \td b^3\, \td g^2}\,,\nn\\
{\td g}' &=& -\, \fft{\td c\, (1-\td g^2)}{4\td a\, \td b\, \td g}\,.\nn
\eea
Note that unlike (\ref{5fo}), the equations in terms of these
variables no longer involve any square roots.  The conserved 
quantities $p$ and $q$ defined in (\ref{pqapp}) are given in terms of
the entirely tilded variables by
\be
p=[(\td a^2-\td b^2)\, \td c - 2\td a\, \td b\, \td f\, \td g_3] \, 
\td g^2\,,\quad
q= -(\td a^2-\td b^2)\, \td c + 2\td a\, \td b\, \td f\,.
\ee

   In the parameterisation of (\ref{d7ans2}), we find that, with
respect to the vielbein basis
\bea
&&\td e^0=dt\,,\quad \td e^1=\td a\, (\Sigma_1 + \td g\,
\sigma_1)\,,\quad \td e^2=\td a\, (\Sigma_2 + \td g\,
\sigma_2)\,,\quad \td e^3 =\td c\, (\Sigma_3 - \sigma_3)\,,\nn\\
&&\td e^4 = \td b\, (\Sigma_1 - \td g\, \sigma_1)\,,\quad
\td e^5 = \td b\, (\Sigma_2 - \td g\, \sigma_2)\,,\quad
\td e^6 = \td f\, (\Sigma_3 + \td g_3\, \sigma_3)\,,\label{viel2}
\eea
(which we denote by $\td e^a$ to distinguish it from the basis $e^a$
in (\ref{viel1})), and the natural choice of spin frame, the
covariantly-constant spinor has purely constant components;
\be
\eta = \fft{\ep_1 + \im\, \ep_2}{\sqrt2}\,,
\ee
where $\ep_1$ and $\ep_2$ are the two constant-component spinors
defined in (\ref{epcon}).  The calibrating 3-form is now given simply by 
\bea
\Phi &=& \td e^0\wedge (\td e^1\wedge \td e^4 + 
\td e^2\wedge \td e^5 + \td e^3\wedge \td e^6)
-(\td e^1\wedge \td e^2-\td e^4\wedge \td e^5)\wedge \td e^3 \nn\\
&&+ (\td e^1\wedge \td e^5-\td e^2\wedge \td e^4)\wedge \td e^6
\,,\label{3form2}
\eea
Of course the first-order equations (\ref{5fo2}) are implied by
$d\Phi=0$ and $d{*\Phi}=0$.  It is interesting that in the metric
parameterisation (\ref{d7ans2}) with the vielbein (\ref{viel2}),
$\Phi$ takes a ``canonical'' form with constant tangent-space
components.

\section{Deformed and resolved conifolds as weak-coupling limits}

  The system of five first-order equations that we have obtained here
encompasses both the four-function first-order system first found in
\cite{bragogugu}, and also the inequivalent system of four first-order
equations plus two constraints that we obtained in \cite{cglpcon}.  To
see the reduction of (\ref{5fo}) and (\ref{constr}) to give this
latter case, it is merely necessary to impose the following further
{\it consistent} algebraic constraint
\be
g=-\fft{b\, \td f}{a\, \td c}\,,\label{alg}
\ee
The metric functions then satisfy a reduced system of first-order
equations, which is most conveniently expressed in terms of the
untilded variables in the form (\ref{d7ans0}) for the metric ansatz:
\bea
\dot a &=& -\fft{c}{2a} + \fft{a^5\, f^2}{8b^4\, c^3}\,,\qquad
\qquad\qquad\quad\
\dot b = -\fft{c}{2b} - \fft{a^2\, (a^2-3c^2)\, f^2}{8b^3\, c^3}\,,\nn\\
\dot c &=& -1 +\fft{c^2}{2a^2} + \fft{c^2}{2b^2}
-\fft{3 a^2\, f^2}{8b^4}\,,\qquad
\dot f = -\fft{a^4\, f^3}{4b^4\,  c^3}\,.\label{newfo}
\eea
These are the first-order equations that were obtained in
\cite{cglpcon}.  The conserved quantities $p$ and $q$ in
(\ref{integrals}) are given by $p=f\, (a^4\, f^2 + 4 b^4\, c^2 -4
a^2\, b^2\, c^2)/(4b^2\, c^2)$ and $q=0$.  The first-order equations
for the resolved conifold emerge from (\ref{newfo}) as the
Gromov-Hausdorff limit, by setting $f\rightarrow 0$.

    To see how the first-order equations (\ref{5fo}) and
(\ref{constr}) can instead be reduced to give the four-function system
in \cite{bragogugu}, one has to impose a different additional
algebraic constraint.  This can be read off by comparing the 
six-function metric ansatz (\ref{d7ans}) with the
four-function ansatz of \cite{bragogugu}, which is
\be
ds_7^2 = dt^2 + a_1^2\, [(\Sigma_1-\sigma_1)^2 + (\Sigma_2-\sigma_2)^2
] + a_3^2\, (\Sigma_3-\sigma_3)^2 + b_1^2 \, [(\Sigma_1+\sigma_1)^2 +
(\Sigma_2+\sigma_2)^2] + b_3^2\, (\Sigma_3+\sigma_3)^2\,.\label{branmet}
\ee
Thus one has
\be
a_1^2=\ft12 a^2\, (1-g)\,,\quad b_1^2=\ft12 a^2\, (1+g)\,,\quad
a_3=\td c\,,\quad b_3=\td f\,,\quad \td g_3 =1\,.
\label{brancon}
\ee
The first-order system (\ref{5fo}) then
consistently reduces to that of \cite{bragogugu}, which is mostly 
simply expressed after changing to the variables $(a_1,b_1,a_3,b_3)$:
\bea
\dot a_1 &=& -\fft{a_1^2}{4a_3\, b_1} + \fft{a_3}{4b_1} +
\fft{b_1}{4a_3} + \fft{b_3}{4a_1}\,,\qquad
\dot a_3 = -\fft{a_3^2}{2a_1\, b_1} + \fft{a_1}{2b_1} + \fft{b_1}{2a_1}
\,,\nn\\
\dot b_1 &=& -\fft{b_1^2}{4a_1\, a_3} +\fft{a_1}{4a_3} +
\fft{a_3}{4a_1} - \fft{b_3}{4b_1}\,,\qquad
\dot b_3 = -\fft{b_3^2}{4a_1^2} + \fft{b_3^2}{4b_1^2}\,.\label{branfo}
\eea
The conserved quantities $p$ and $q$ defined in (\ref{integrals}) are
given by $p=-q=a_3\, (b_1^2-a_1^2) + 2a_1\, b_1\, b_3$.  By taking the
limit $b_3\longrightarrow0$ in (\ref{branfo}), the equations reduce to
those that describe the deformed conifold in $D=6$.

   Note that the more symmetrical version of the metric (\ref{d7ans2}) 
is especially suited to taking the Gromov-Hausdorff limit to the 
deformed conifold.  By comparing (\ref{d7ans2}) and (\ref{branmet}),
we see that the constraint that reduces (\ref{d7ans2}) to
(\ref{branmet}) is now simply $\td g=-1$. 

  The first-order system (\ref{branfo}) gives rise to the $\bB_7$
solutions, whose Gromov-Hausdorff limit is $S^1$ times the deformed
conifold, and to the $\bC_7$ solutions, whose Gromov-Hausdorff limit
is $S^1$ times the K\"ahler metric on the complex line bundle over
$S^2\times S^2$ obtained in \cite{berber,pagpop}.  These both have
$p=-q$, with $p=-a_0^3$ for $\bB_7$, where $a_0$ is the radius of the
$S^3$ bolt, and $p=a_0^2\, c_0$ for $\bC_7$, where $a_0$ and $c_0$ are
the radii of $S^2$ in the $S^2\times S^2$ base and the $U(1)$ fibre of
the $T^{1,1}$ bolt.  By contrast, the first-order system obtained in
\cite{cglpcon} gives rise to the $\bD_7$ solutions, with $q=0$ and $p$
proportional to the volume of the squashed $S^3$ bolt.  The
Gromov-Hausdorff limit of the $\bD_7$ metrics is the resolved
conifold.

\section{New $G_2$ metrics $\wtd \bC_7$ with $T^{1,1}$ bolt}

    We find that as well as yielding all the above $\bB_7$, $\bC_7$
and $\bD_7$ solutions, the full system with five first-order equations
that we have obtained in this paper gives rise to another new class of
complete non-singular $G_2$ metrics, which we shall denote by
$\wtd\bC_7$, with two non-trivial parameters.  At an upper boundary of
the parameter range the metrics are AC, whilst away from this boundary
they are ALC.  At a lower boundary of the parameter range the
Gromov-Hausdorff limit is reached, where the $\wtd\bC_7$ metrics
approach $S^1$ times the Ricci-flat K\"ahler 6-metrics obtained in
\cite{pando2}.  These 6-metrics, which we shall denote by $\wtd
\bK_6$, have the topology of a complex line bundle over $S^2\times
S^2$, and they are generalisations of the example found in
\cite{berber,pagpop}.  Specifically, the radii of the two 2-spheres in
the $S^2\times S^2$ bolt in the metrics of \cite{pando2} can be chosen
arbitrarily, with the metric in \cite{berber,pagpop}, which we shall
denote by $\bK_6$, being the special case where the two radii are
equal.  Thus the $\wtd\bC_7$ metrics with the radii of the two $S^2$
factors chosen to be equal are precisely the $\bC_7$ metrics found in
\cite{slumpy}.

   To obtain these new solutions $\wtd\bC_7$, we begin by constructing
regular short-distance expansions in the form of Taylor series.
Substituting into (\ref{5fo}) and (\ref{constr}), we find that the
metric functions at short distance take the form
\bea
a&=& a_0 + \fft{(4 a_0^2- \td f_0^2)\, t^2}{16 a_0^3} + \cdots\,,\nn\\
b&=& b_0 + \fft{(4 a_0^4 - 3 b_0^2\, \td f_0^2)\, t^2}{16 b_0\, a_0^4} +
\cdots\,,\nn\\
c &=& -t + \fft{(4 a_0^2\, b_0^2 +4 a_0^4 - 
            b_0^2\, \td f_0^2)\, t^2}{24 a_0^4} + \cdots\,,\nn\\
\td f &=& \td f_0 + \fft{\td f_0^3\, t^2}{4a_0^4}  +\cdots\,,\nn\\
g&=& \fft{b_0\, \td f_0\, t }{2a_0^3} + \fft{\td f_0\, (4 a_0^4 - 20 a_0^2\,
b_0^2 + 11 b_0^2\, \td f_0^2)\, t^3}{96 a_0^7\, b_0} + \cdots\,,\nn\\
\td g_3 &=& \fft{b_0^2}{a_0^2} + \fft{(a_0^4 -b_0^4)\, t^4}{8 a_0^6\, b_0^2} 
+\cdots\,,\label{taylorexp}
\eea
where $a_0$, $b_0$ and $\td f_0$ are free parameters.  One of the three
parameters here can be viewed as being trivial, since it is just
associated with the overall scale of the solution.  Two out of the
three, for example $\td f_0/a_0$ and $\td f_0/b_0$, constitute non-trivial
parameters in the solutions.  They characterise the homogeneous
squashings of the $T^{1,1}$ bolt at $t=0$.

   By using the Taylor expansions to set initial data just outside the
bolt, and then integrating the first-order equations numerically, we
find that the solutions are regular also at large distance, provided
that the parameters $a_0$, $b_0$ and $\td f_0$ lie in appropriate
ranges.  There is a two-dimensional non-trivial parameter space of
regular solutions $\wtd\bC_7$.  For generic points in this modulus
space, the metric is ALC, with a $T^{1,1}$ bolt whose squashing is
characterised by $a_0$, $b_0$ and $c_0$.  At large distance, the
metric approaches a twisted product of $S^1$ and an AC six-metric,
namely the Ricci-flat K\"ahler metric on the $\R^2$ bundle over
$S^2\times S^2$ with, generically, unequal $S^2$ radii.  There is a
boundary of the modulus space corresponding to the situation where the
radius of this circle goes to infinity, and the seven-metric becomes
AC.  As in the special case $\bC_7$ when $a_0=b_0$ \cite{slumpy}, the
magnitude of the Killing vector $K$ given in (\ref{kvec}) is
everywhere finite and non-zero in the ALC $\wtd \bC_7$ metrics.  It is
given by $|K|^2 = \td f^2\, (1+\td g_3)^2$, and this runs from a
minimum value at $t=0$, where from (\ref{taylorexp}) we have
$|K|_{t=0}= \td f_0\, (1+ b_0^2/a_0^2)=\td f_0\, (1+\sqrt{-p/q})$, to a
final value $|K|_{t=\infty} =2 \td f_\infty$ (since $\td g_3$ goes to
1 at infinity), which stabilises at infinity.  In the Gromov-Hausdorff
limit the $\wtd\bC_7$ metrics become $S^1$ times the Ricci-flat
K\"ahler metrics found in \cite{pando2}.

    It is straightforward to see, by substituting the Taylor 
expansions (\ref{taylorexp}) into the definitions of the first integrals
$p$ and $q$ given in (\ref{integrals}), that we have
\be
p= -\, \fft{b_0^4\, \td f_0}{a_0^2}\,,\qquad q = a_0^2\, \td f_0\,.
\ee
Thus the conserved quantities $p$ and $q$ characterise the relative
sizes of the two $S^2$ factors in the $S^2\times S^2$ base space of
the $T^{1,1}$ bolt.  Note, in particular, that if we set the two radii
equal, $a_0=b_0$, then we shall have $p=-q= -a_0^2\, \td f_0$.  This special
case, which corresponds to the $\bC_7$ metrics found in \cite{slumpy},
has the $Z_2$ symmetry $p=-q$ under the interchange of the two $S^3$
factors in the $S^3\times S^3$ principal orbits.  The $\wtd \bC_7$
metrics that we have found here have $p\ne -q$, and hence they are not
$Z_2$ symmetric.  Note that it is because of the geometry of the
$\wtd \bC_7$ manifolds, with the $T^{1,1}$ bolt as opposed to the $S^3$
bolt of the $\bB_7$ and $\bD_7$ manifolds, that we can obtain smooth
$G_2$ metrics with $p$ and $q$ both non-zero and $p\ne -q$.  In fact
this is the only possible geometry for smooth metrics with such $p$
and $q$ values.

\section{Discussion}

    The $G_2$ metrics can be used in order to construct supersymmetric
Ricci-flat solutions of M-theory, by simply writing $d\hat s_{11}^2 = 
dx^\mu\, dx_\mu + ds_7^2$.  There is a $U(1)$ isometry, and so we can
reduce this solution to $D=10$, using the standard Kaluza-Klein
formula
\be
d\hat s_{11}^2 = e^{-\fft16 \varphi}\, ds_{10}^2 + e^{\fft43\varphi}\, 
(dz+ A)^2\,.\label{mto2a}
\ee
Since the $U(1)$ Killing vector is given by (\ref{kvec}), we shall
have $z=\ft12(\psi + \wtd\psi)$ as the fibre coordinate, together with
$y=\ft12(\psi-\wtd\psi)$ as a coordinate in $ds_{10}^2$.  The $G_2$ metric
(\ref{d7ans}), or (\ref{d7ans2}), which has the form
$ds_7^2=ds_6^2 + \td f^2\, (\Sigma_3 + \td g_3\, \sigma_3)^2$, can now
be written as
\be
ds_7^2 = ds_6^2 + \td f^2\, (1+\td g_3)^2\, (dz + A)^2\,,
\ee
where the Kaluza-Klein vector is given by
\be
A= \fft1{1+\td g_3}\, \Big[ \td g_3\, (dy+ \cos\theta\, d\phi) - 
               (dy-\cos\td\theta\, d\td\phi)\Big]\,.
\ee
The $D=10$ configuration therefore carries non-trivial Ramond-Ramond 2-form
flux. The functions $\td f$ and $(1+\td g_3)$ are everywhere finite and  
non-vanishing in the $\bC_7$ \cite{slumpy} and 
$\bD_7$ \cite{cglpcon} solutions, and also in the $\wtd\bC_7$
solutions that we have found in this paper.   This implies that in
the reduction (\ref{mto2a}) to the type IIA theory, the string
coupling $g_{\rm string} = e^{\varphi}= \td f^{3/2}\, (1+\td g_3)^{3/2}$ 
is everywhere finite and non-zero.  This may provide a good supergravity
dual for ${\cal N}=1$ field theory in four dimensions.

   It is useful to summarise the results for known classes of $G_2$
metrics with $S^3\times S^3$ principal orbits in a Table. 
\bigskip\bigskip

\centerline{
\begin{tabular}{|c|c|c|c|}\hline
$G_2$ Metric & Gromov-Hausdorff Limit  & Bolt & $(p,q)$ \\ \hline\hline
$\bB_7$ & Deformed conifold & $S^3_1$
& $p=-q$  \\ \hline
$\bC_7$  & $\bK_6$
& $T^{1,1}_\lambda$ & $p=-q$ \\ \hline
$\wtd \bC_7$  & $\wtd \bK_6$
& $T^{1,1}_{\lambda,\mu}$ & $p\ne -q$ \\ \hline
$\bD_7$ & Resolved conifold & $S^3_\lambda$
& $q=0$ \\ \hline
AC$_7$ & -- & $S^3_1$ & $(p,q)=(p,0), (p,-p), (0,q)$\\ \hline
\end{tabular}}
\bigskip

\centerline{Table 1: The $G_2$ metrics with $S^3\times S^3$ principal orbits}
\bigskip\bigskip

   The space denoted by AC$_7$ is the original AC $G_2$ metric found
in \cite{brysal,gibpagpop}.  The spaces denoted by $\bK_6$ and
$\wtd\bK_6$ are the Ricci-flat K\"ahler metrics on the complex line
bundle over the Einstein $S^2\times S^2$ base \cite{berber,pagpop},
and its generalisation found in \cite{pando2}, respectively.  The
subscripts $\lambda$ and $\mu$ on the spaces $S^3_\lambda$,
$T^{1,1}_\lambda$ and $T^{1,1}_{\lambda,\mu}$ denote the adjustable
parameters that characterise the homogeneous squashings of the bolt.
There are three possible choices for the relation between $p$ and $q$
in the original AC metric AC$_7$, since there is a triality symmetry
in this case \cite{atiwit,cglp3,brand}.

   The most general solution of our system of five first-order
equations for $G_2$ holonomy would have five modulus parameters,
corresponding to the constants of integration of the equations.  Two
of these are trivial, corresponding to the freedom to redefine the
radial coordinate by the addition and multiplication by constants.
(The latter changes the overall scale of the metric.)  Three
non-trivial parameters remain, but this solution would in general be
singular.  All the regular solutions, listed in Table 1, arise in special
regions of the modulus space.  The original AC$_7$ metric of
\cite{brysal,gibpagpop} has no non-trivial parameter; the $\bB_7$ and
$\bD_7$ metrics have one non-trivial parameter; and the new
$\wtd\bC_7$ metrics have two non-trivial parameters ($\bC_7$ is a
special case of these with one non-trivial parameter). Thus all of the
metrics in Table 1, and their weak-coupling limits, are unified within 
the general $G_2$ metrics described by the five first-order equations,
provided one includes the singular metrics that interconnect them.

\section*{Acknowledgements}

   G.W.G., H.L. and C.N.P. are grateful to UPenn for support and
hospitality during the course of this work.

\end{document}